\documentstyle[preprint,prd,aps,epsf]{revtex}

\begin{document}
\draft
\title{\bf
Dyons as a source of $CP$ and
time invariance violation: electric dipole moments
and K-meson decays}
\author{D. W. Murray and V. V. Flambaum}
\address{School of Physics, University of New South Wales,
Sydney, 2052, Australia}
%\date{\today}
\maketitle
\begin{abstract}
We consider a mechanism by which dyons (electrically charged
magnetic monopoles) can produce both a $T$- and $P$-odd
(i.e. time reversal invariance and parity violating)
mixed polarizability $\beta$ [defined
by $\Delta E = -\beta {\bf E} \cdot {\bf B}$, where
$\Delta E$ is the energy change when electric (${\bf E}$)
and magnetic (${\bf B}$) fields are applied to a system]
and a $T$- and $P$-odd interaction
between two particles:
$\overline{\psi}_1 \gamma_5 \psi_1 \overline{\psi}_2
\psi_2$, where the $\psi_i$ are electron and quark spinors.
The latter can create
atomic and neutron electric dipole moments (EDMs).
From experimental
bounds on these we find limits on the properties
of dyons. Our best limit, using the
experimental limit for the EDM of the Tl atom, is
$M |Q g (Q^2 - g^2)|^{-1/4} > 6 \mbox{ GeV}$, where
$M$ is the dyon mass and $Q$ is the electric and
$g$ the magnetic charge of the dyons. The
contribution of dyons to $CP$ violation in
K-meson decays is also estimated.
\end{abstract}
\vspace{5mm}
\vspace{10mm}

\section{Introduction}
\label{sint}
It would be very interesting if $CP$ and $T$ violation
were a consequence of another fundamental symmetry,
such as a symmetry between electric and magnetic charges.
In Ref. \cite{FM97} we considered the possibility
that dyons (electrically charged magnetic monopoles)
could induce parity and time invariance
violating electric dipole moments (EDMs) of
quantum systems.
The point is that the polarization of the dyon vacuum by
the Coulomb field may produce not only the usual correction
to the electric field, $\delta {\bf E}$, but also a radial
magnetic field ${\bf B} = (g/Q) \delta {\bf E}$, where
$Q$ is the electric and $g$ the magnetic charge of the dyons
(the corresponding antidyons would have charges $-Q$ and $-g$).
The interaction of electrons (or quarks) with this magnetic
field could generate atomic (and neutron) electric dipole
moments.
[In
addition, electron and
neutron EDMs $\propto m/M^2$ could
appear due to higher order diagrams ($m$ is the electron or
quark mass; $M$ is the dyon mass).]
The EDMs were found to be proportional to $Q g$.
If there were a second dyon-antidyon pair, with charges
$Q$ and $-g$, and $-Q$ and $g$, then they would also make
a contribution, which would
cancel out the original contribution (assuming this second pair
had the same mass as the first). Therefore to have
time invariance violation we can only have one
of the dyon-antidyon pairs.

Our consideration in
\cite{FM97} gave quite a strong limit on the dyon mass:
$M / \sqrt{|Qg|} > 100 \mbox{ TeV}$.
However, this consideration was only based on
``heuristic'' arguments; it did not prove that the
effect exists, only that it is conceivable. 

In Ref. \cite{KOST97} a different mechanism was considered,
by which
dyons may induce an electron EDM. Firstly, they calculated the
lowest order $T$- and $P$-odd correction to the effective
Lagrangian of QED due to dyon vacuum polarization. This
involves finding the change in the energy density of the
vacuum dyons when external electric and magnetic fields are
applied. The correction $\Delta L_d$ was found to be
\begin{equation}
\Delta L_d \approx \frac{Q g (Q^2 - g^2)}{60 \pi^2 M^4}
({\bf B} \cdot {\bf E}) ({\bf B}^2 - {\bf E}^2).
\label{epp2b}
\end{equation}
for (spin $1/2$) dyons
with masses $M$, electric charges $Q$,
and magnetic charges $g$.
Under time reversal ($t \rightarrow -t$) and 
parity (${\bf r}
\rightarrow -{\bf r}$)
transformations ${\bf B} \cdot {\bf E}$ changes
sign, while ${\bf B}^2 - {\bf E}^2$
is unchanged, and so $\Delta L_d$ is
$T$- and $P$-odd.

In Ref. \cite{KOST97} they used this effective Lagrangian
to estimate the induced electron EDM. Unfortunately,
the integrals involved in such calculations are quadratically
divergent. Therefore, the calculation cannot be done within
the limits of the applicability of
the Lagrangian (\ref{epp2b}).
The result depends strongly on the effective cut-off parameter.

In this paper we present further arguments that dyon
vacuum polarization may produce $T$ and $P$ violating effects.
As in Ref. \cite{KOST97}, our calculation is based on the
effective Lagrangian $\Delta L_d$ (\ref{epp2b}). However,
the integrals involved in the calculation of the effects
considered here (mixed polarizablity, $T$- and $P$-odd
scalar-pseudoscalar interaction between electrons and quarks,
and atomic and neutron EDMs) are only
logarithmically divergent.
Therefore, we stay within the limit of the applicability of
the effective Lagrangian (\ref{epp2b}).

\begin{figure}
\leavevmode
\begin{center}
\begin{tabular}{c}
\epsfxsize=245pt
\epsffile{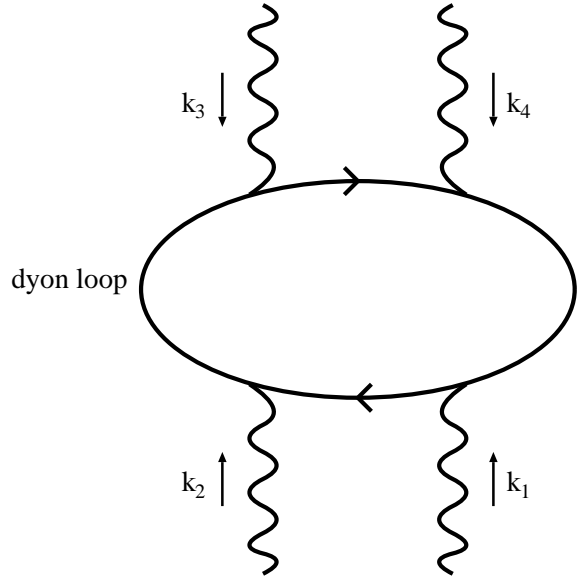}
\end{tabular}
\end{center}
\caption{Diagram showing photon-photon scattering due to
dyon vacuum polarization. We take all photon momenta as
pointing inward.}
\label{f1}
\end{figure}

In \cite{KOST97} the photon-photon scattering tensor
(see, e.g., \cite{LL2}, \S 124 and \cite{AkhBer}, \S 54.4 for
the definition of this) for low-energy photon-photon scattering
due to dyons was
derived from $\Delta L_d$ (this
is the basis for our calculation
of the dyon effects).
Photon-photon scattering due to dyon vacuum polarization is
shown in Fig. \ref{f1}. The result for this $T$- and $P$-odd
photon-photon scattering tensor is
\begin{eqnarray}
M^{\mu \nu \rho \sigma} (k_1,k_2,k_3,k_4) & = &
\frac{2 Q g (Q^2 - g^2)}{45 M^4}
[{\varepsilon_{\alpha \beta}}^{\mu \nu}
k_1^\alpha k_2^\beta k_3^\sigma k_4^\rho
+ {\varepsilon_{\alpha \beta}}^{\mu \rho}
k_1^\alpha k_2^\sigma k_3^\beta k_4^\nu
+ {\varepsilon_{\alpha \beta}}^{\mu \sigma}
k_1^\alpha k_2^\rho k_3^\nu k_4^\beta
\nonumber \\
& & \mbox{} + {\varepsilon_{\alpha \beta}}^{\nu \rho}
k_1^\sigma k_2^\alpha k_3^\beta k_4^\mu
+ {\varepsilon_{\alpha \beta}}^{\nu \sigma}
k_1^\rho k_2^\alpha k_3^\mu k_4^\beta
+ {\varepsilon_{\alpha \beta}}^{\rho \sigma}
k_1^\nu k_2^\mu k_3^\alpha k_4^\beta
\nonumber \\
& & \mbox{} - {\varepsilon_{\alpha \beta}}^{\mu \nu}
g^{\rho \sigma}
(k_3 k_4) k_1^\alpha k_2^\beta
- {\varepsilon_{\alpha \beta}}^{\mu \rho} g^{\nu \sigma}
(k_2 k_4) k_1^\alpha k_3^\beta
- {\varepsilon_{\alpha \beta}}^{\mu \sigma} g^{\rho \nu}
(k_2 k_3) k_1^\alpha k_4^\beta
\nonumber \\
& & \mbox{}
- {\varepsilon_{\alpha \beta}}^{\nu \rho} g^{\mu \sigma}
(k_1 k_4) k_2^\alpha k_3^\beta
- {\varepsilon_{\alpha \beta}}^{\nu \sigma} g^{\mu \rho}
(k_1 k_3) k_2^\alpha k_4^\beta
- {\varepsilon_{\alpha \beta}}^{\rho \sigma} g^{\mu \nu}
(k_1 k_2) k_3^\alpha k_4^\beta]
\label{eppst}
\end{eqnarray}

In Sec. \ref{smixpol} we find the dyon
induced $T$- and $P$-odd mixed polarizability of a particle, as
well as a nucleus. The $T$- and $P$-odd mixed polarizability,
$\beta$, can be defined by the equation $\Delta E = - \beta
{\bf E} \cdot {\bf B}$, where $\Delta E$ is the
change in energy
of a system when electric (${\bf E}$) and magnetic (${\bf B}$)
fields are applied. In Sec. \ref{stpoint} we find the
$T$- and $P$-odd interaction between two particles that could
be induced by dyons and in Sec. \ref{sdexp} we use experimental
bounds on EDMs to give limits on dyon properties.
Finally, in Sec. \ref{skmeson} we estimate the contribution
of dyons to $CP$ violation in K-meson decays.

Note that it is possible that dyons do not exist as free
particles but only as bound dyon-antidyon pairs that cannot
be separated (for example, Nambu showed that in
the standard electroweak model the classical solution is a
monopole-antimonopole pair connected by a $Z^0$ field string
\cite{Nambu1977}). In this case dyons with a small mass are
not ruled out by experimental searches for
new particles (a bound state of dyons may be hard to produce
and identify). The effects considered in this paper
apply equally well for free dyons and bound dyons.

\section{Dyon induced $T$- and $P$-odd mixed polarizability}
\label{smixpol}
\begin{figure}
\leavevmode
\begin{center}
\begin{tabular}{c}
\epsfxsize=245pt
\epsffile{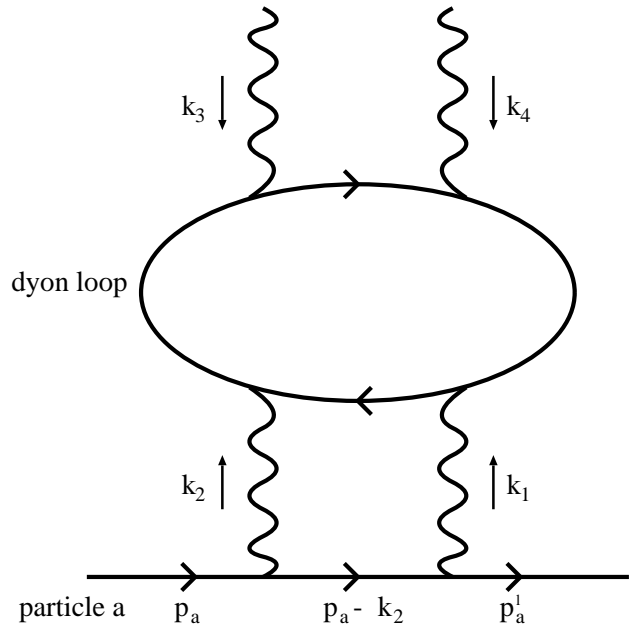}
\end{tabular}
\end{center}
\caption{Diagram showing the dyon 
induced $T$- and $P$-odd mixed
polarizability.}
\label{f2}
\end{figure}

We will now find the matrix element
corresponding to Fig. \ref{f2} using the photon-photon
scattering tensor (\ref{eppst}).
Note that because of the way in which the photon-photon
scattering tensor was symmetrized in \cite{KOST97} we are
actually calculating the matrix element for all of the
diagrams involving different arrangements of the photons
(e.g. the diagram similar to Fig. \ref{f2}, but with
the lower two photons crossing each other) put together.
This only covers diagrams with two photons free and
two attached to a particle (or, when two particles
are involved, as in Fig. \ref{f3},
two photons attached to each particle).
Other diagrams (such as that considered in \cite{KOST97})
do not involve logarithmically divergent integrals ---
our aim is not to do a calculation including every
type of diagram with dyons, rather it is to show that
the effect exists.

Using the Feynman rules we
get (we use $e^2 = \alpha$ and the Feynman rules
and notation of Ref. \cite{LL2})
\begin{eqnarray}
M_{fi}
& = & \frac{i e_a^2}{(2 \pi)^4 (4 \pi)^2} A_\lambda (k_3)
A_\omega(k_4) \nonumber \\
& & \times
\int \overline{u} (p_a^{\prime}) \gamma^\mu G(p_a-k_2)
\gamma^\nu u(p_a)
D_{\alpha \mu} (k_1) D_{\beta \nu} (k_2) 
M^{\alpha \beta \lambda
\omega} (k_1,k_2,k_3,k_4) \, d^4 k_2,
\label{ea1}
\end{eqnarray}
where $G(k)$ and $D_{\alpha \beta} (k)$ are the fermion and
photon propagators,
$e_a$ is the charge and $m_a$ the mass of particle $a$,
and we use the notation
$\hat{p} \equiv p \! \! \! / \equiv p^\mu \gamma_\mu$.
We define $q$ as $q = p_a - p_a^\prime$, and so we have
$k_1 = -k_2 + q$. Substituting in the propagators and
simplifying gives
\begin{eqnarray}
M_{fi} 
& = & \frac{i e_a^2}{(2 \pi)^4} A_\lambda (k_3) A_\omega (k_4)
\int \overline{u} (p_a^\prime) 
\gamma_\mu \frac{\hat{p_a}-\hat{k_2}+m_a}
{(p_a-k_2)^2 - m_a^2 + i \epsilon} 
\gamma_\nu u(p_a) \nonumber \\
& & \times
\frac{1}{(k_2^2 + i \epsilon) [(k_2 - q)^2 + i \epsilon]}
M^{\mu \nu \lambda \omega} (k_1,k_2,k_3,k_4) \, d^4 k_2.
\label{ea2}
\end{eqnarray}
We now use the following identity, which can be obtained in a
similar way to Eq. (24) of Ref. \cite{KOST97}:
\begin{equation}
M^{\mu \nu \lambda \omega} (k_1,k_2,k_3,k_4)
= k_{3 \rho} k_{4 \sigma} 
\frac{\partial}{\partial k_{3 \lambda}}
\frac{\partial}{\partial k_{4 \omega}} M^{\mu \nu \rho \sigma}
(k_1,k_2,k_3,k_4).
\label{ea3}
\end{equation}
This gives
\begin{eqnarray}
M_{fi} & = & \frac{i e_a^2}{(2 \pi)^4} 
A_\lambda (k_3) A_\omega (k_4)
k_{3 \rho} k_{4 \sigma} \nonumber \\
& & \times \overline{u} (p_a^\prime) \gamma_\mu
\left[ \int \frac{\hat{p_a} - \hat{k_2} + m_a}{[(p_a-k_2)^2
- m_a^2 + i \epsilon]
(k_2^2 + i \epsilon) [(k_2 - q)^2 + i \epsilon]}
\frac{\partial}{\partial k_{3 \lambda}} 
\frac{\partial}{\partial
k_{4 \omega}} M^{\mu \nu \rho \sigma} \, d^4 k_2 \right] 
\gamma_\nu u(p_a),
\label{ea4}
\end{eqnarray}
where
\begin{eqnarray}
\frac{\partial}{\partial k_{3 \lambda}} \frac{\partial}
{\partial k_{4 \omega}} 
M^{\mu \nu \rho \sigma} (k_1,k_2,k_3,k_4)
& = & \frac{2 Q g (Q^2 - g^2)}{45 M^4}
[g^{\lambda \sigma} g^{\omega \rho} 
{\varepsilon_{\alpha \beta}}^{\mu
\nu} k_1^\alpha k_2^\beta
+ g^{\omega \nu} {\varepsilon_{\alpha}}^{\lambda \mu \rho}
k_1^\alpha k_2^\sigma
+ g^{\lambda \nu} {\varepsilon_{\alpha}}^{\omega \mu \sigma}
k_1^\alpha k_2^\rho
\nonumber \\
& & \mbox{} 
+ g^{\omega \mu} {\varepsilon_{\alpha}}^{\lambda \nu \rho}
k_1^\sigma k_2^\alpha
+ g^{\lambda \mu} {\varepsilon_{\alpha}}^{\omega \nu \sigma}
k_1^\rho k_2^\alpha
+ \varepsilon^{\lambda \omega \rho \sigma}
k_1^\nu k_2^\mu
\nonumber \\
& & \mbox{} 
- g^{\lambda \omega} g^{\rho \sigma} 
{\varepsilon_{\alpha \beta}}
^{\mu \nu} k_1^\alpha k_2^\beta
- g^{\nu \sigma} {\varepsilon_\alpha}^{\lambda \mu \rho}
k_1^\alpha k_2^\omega
- g^{\rho \nu} {\varepsilon_\alpha}^{\omega \mu \sigma}
k_1^\alpha k_2^\lambda
\nonumber \\
& & \mbox{}
- g^{\mu \sigma} {\varepsilon_\alpha}^{\lambda \nu \rho}
k_1^\omega k_2^\alpha
- g^{\mu \rho} {\varepsilon_\alpha}^{\omega \nu \sigma}
k_1^\lambda k_2^\alpha
- g^{\mu \nu} \varepsilon^{\lambda \omega \rho \sigma}
k_1^\pi k_{2 \pi}].
\label{ea5}
\end{eqnarray}
Using this equation (and keeping in mind that
$k_1 = q - k_2$), we can see that the integral in
Eq. (\ref{ea4}) will be made up of terms that contain the form
\begin{equation}
\int 
\frac{(\hat{p_a} - \hat{k_2} + m_a) (q^\alpha - k_2^\alpha)
k_2^\beta \, d^4 k_2}{(k_2^2 + i \epsilon) [(k_2 - q)^2 +
i \epsilon] [(k_2 - p_a)^2 - m_a^2 + i \epsilon]}
\equiv W^{\alpha \beta},
\label{ea6}
\end{equation}
which we define as $W^{\alpha \beta}$ for convenience
($\alpha$ and $\beta$ can stand for any indices).
Using this notation we can write
\begin{eqnarray}
M_{fi} & = & \frac{i e_a^2 Q g (Q^2 - g^2)}{360 \pi^4 M^4}
 A_\lambda (k_3) A_\omega (k_4) k_{3 \rho}
k_{4 \sigma}
\nonumber \\
& & \times \overline{u} (p_a^\prime) \gamma_\mu
(
g^{\lambda \sigma} g^{\omega \rho} 
{\varepsilon_{\alpha \beta}}^{\mu
\nu} W^{\alpha \beta}
+ g^{\omega \nu} {\varepsilon_{\alpha}}^{\lambda \mu \rho}
W^{\alpha \sigma}
+ g^{\lambda \nu} {\varepsilon_{\alpha}}^{\omega \mu \sigma}
W^{\alpha \rho}
\nonumber \\
& & \mbox{} 
+ g^{\omega \mu} {\varepsilon_{\alpha}}^{\lambda \nu \rho}
W^{\sigma \alpha}
+ g^{\lambda \mu} {\varepsilon_{\alpha}}^{\omega \nu \sigma}
W^{\rho \alpha}
+ \varepsilon^{\lambda \omega \rho \sigma}
W^{\nu \mu}
\nonumber \\
& & \mbox{} 
- g^{\lambda \omega} g^{\rho \sigma} 
{\varepsilon_{\alpha \beta}}
^{\mu \nu} W^{\alpha \beta}
- g^{\nu \sigma} {\varepsilon_\alpha}^{\lambda \mu \rho}
W^{\alpha \omega}
- g^{\rho \nu} {\varepsilon_\alpha}^{\omega \mu \sigma}
W^{\alpha \lambda}
\nonumber \\
& & \mbox{}
- g^{\mu \sigma} {\varepsilon_\alpha}^{\lambda \nu \rho}
W^{\omega \alpha}
- g^{\mu \rho} {\varepsilon_\alpha}^{\omega \nu \sigma}
W^{\lambda \alpha}
- g^{\mu \nu} \varepsilon^{\lambda \omega \rho \sigma}
{W^{\pi}}_\pi
)
\gamma_\nu u(p_a)
\label{ea14}
\end{eqnarray}

$W^{\alpha \beta}$ can be worked out using the Feynman
parameterization technique. We use the identity (see, e.g.,
\cite{LL2,MSQFT})
\begin{equation}
\frac{1}{abc} = 2 \int_0^1 dx \!\! \int_0^{1-x} dz
\frac{1}{[a + (b-a) x + (c-a) z]^3},
\label{ea7}
\end{equation}
with $a = k_2^2$, $b = (k_2 - q)^2$, and
$c = (k_2 - p_a)^2 - m_a^2$ (we now omit the
$i \epsilon$'s). Using this, and completing the square in the
denominator gives
\begin{equation}
W^{\alpha \beta} = 2 \int_0^1 dx \!\! \int_0^{1-x} dz
\!\! \int \frac{(\hat{p_a} - \hat{k_2} + m_a) (q^\alpha
k_2^\beta - k_2^\alpha k_2^\beta)}{[(k_2 - l)^2 - t^2]^3}
\, d^4 k_2,
\label{ea8}
\end{equation}
where
\begin{equation}
l = x q + z p_a,
\label{ea9}
\end{equation}
and
\begin{equation}
t^2 = (x q + z p_a)^2 - x q^2 - z p_a^2 + z m_a^2.
\label{ea10}
\end{equation}
We now do a change of variable:
$k_2 \rightarrow k_2 + l$.
The numerator in the integral in Eq. (\ref{ea8}) changes
as follows:
\begin{eqnarray}
(\hat{p_a} - \hat{k_2} + m_a) (q^\alpha k_2^\beta
- k_2^\alpha k_2^\beta)
& \rightarrow & [-m_a + (z-1) \hat{p_a} + x \hat{q}] 
k_2^\alpha
k_2^\beta
+ z (p_a^\alpha k_2^\beta \hat{k_2}
+ p_a^\beta k_2^\alpha \hat{k_2})
\nonumber \\
& &
\mbox{} + x (q^\alpha k_2^\beta \hat{k_2} + q^\beta k_2^\alpha
\hat{k_2}) - q^\alpha k_2^\beta \hat{k_2},
\label{ea11}
\end{eqnarray}
where we threw away the terms that are odd in $k_2$, as they 
will
vanish on integration, as well as the terms independent 
of $k_2$,
as their contribution will be much smaller than the $\propto
{k_2}^2$ terms, which are logarithmically divergent when 
integrated.
In the integral we can
replace $k_2^\alpha k_2^\beta$ with
$\frac{1}{4} g^{\alpha \beta} k_2^2$
(see, e.g, \cite{LL2}, \S 127).
So
\begin{eqnarray}
W^{\alpha \beta} 
& = & \frac{1}{2} \int_0^1 dx \!\! \int_0^{1-x} dz
\!\! \int \frac{k_2^2 \, d^4 k_2}{(k_2^2 - t^2)^3}
\nonumber \\
& & \times
\{ [ -m_a + (z-1)\hat{p_a} + x \hat{q}] g^{\alpha \beta}
+ z (p_a^\alpha \gamma^\beta + p_a^\beta \gamma^\alpha)
+ x (q^\alpha \gamma^\beta + q^\beta \gamma^\alpha)
- q^\alpha \gamma^\beta \}
\label{ea12}
\end{eqnarray}

Consider the 
integral $\int k_2^2 (k_2^2 - t^2)^{-3} \, d^4 k_2$.
For $k_2 \gg m_a$ the integrand is proportional
to $1 / k_2$ 
(since $d^4 k_2 \propto k_2^3 \, dk_2$) and hence the integral
will be a logarithm. We take a lower ``cut-off'' for this
integral of $k_2 = m_a$, since near this point 
the $\propto 1/k_2$
behavior of the integrand begins to break down. For our
upper ``cut-off'' we take the dyon mass $M$, because, as
stated in \cite{KOST97}, Eq. (\ref{eppst}) is only valid when
the photon momenta are small compared to $M$. Hence we have
$\int k_2^2 (k_2^2 - t^2)^{-3} \, d^4 k_2 \approx 2 \pi^2 i
\ln (M/m_a)$.
(The factor of $i$ appears as $k_2$ is a $4$-vector 
in Minkowski space.)
Using this
and integrating over $x$ and $z$ gives
\begin{equation}
W^{\alpha \beta} = i \frac{\pi^2}{6} 
\ln \left( \frac{M}{m_a} \right)
[ (-3 m_a - 2 \hat{p_a} + \hat{q}) g^{\alpha \beta}
+ (p_a^\alpha \gamma^\beta + p_a^\beta \gamma^\alpha)
+ (q^\alpha \gamma^\beta + q^\beta \gamma^\alpha)
- 3 q^\alpha \gamma^\beta]
\label{ea13}
\end{equation}

We now put this result for $W^{\alpha \beta}$ into
Eq. (\ref{ea14}). Evaluating this is a long and involved
process and so we will not give the details here.
To give an idea of how the calculation is done,
the calculation for one of the terms --- the
term $c_1 \hat{p_a} g^{\alpha \beta}$ in $W^{\alpha \beta}$,
where $c_1 = -i (\pi^2 / 3) \ln (M/m_a)$ --- is done in
the Appendix. We denote the
contribution of this term to $M_{fi}$ by $M_{fi}^{(2)}$.
The result is [see Eq. (\ref{ea35})]
$M_{fi}^{(2)} =
e_a^2 Q g (Q^2 - g^2) / (216 \pi^2 M^4) m_a
\ln (M/m_a) \overline{u} (p_a^\prime) u(p_a) 
\widetilde{F}^{\alpha \beta}
(k_3) F_{\alpha \beta} (k_4)$.

The $m_a g^{\alpha \beta}$ and $(p_a^\alpha \gamma^\beta
+ p_a^\beta \gamma^\alpha)$ terms in $W^{\alpha \beta}$ give
similar structures. The contribution of the
$\hat{q} g^{\alpha \beta}$ and $(q^\alpha \gamma^\beta
+ q^\beta \gamma^\alpha)$ terms to $M_{fi}$ vanishes
--- this is essentially due to the fact that
$\overline{u} (p_a^\prime) \hat{q} u(p_a) = 0$, which is 
just the
charge conservation condition and can
be proved from the Dirac equation 
(using $q = p_a - p_a^\prime$).
Note that all of these terms are symmetric in $\alpha$ and
$\beta$.

The remaining, nonsymmetric term in $W^{\alpha \beta}$,
$\propto -3 q^\alpha \gamma^\beta$ gives a 
different structure:
$i [e_a^2 Q g (Q^2 - g^2)] / (144 \pi^2 M^4) \ln (M/m_a)
\overline{u} (p_a^\prime) \hat{q} \gamma_5 u(p_a) 
F_{\alpha \beta} (k_3)
F^{\alpha \beta} (k_4)$.

Using all the terms of $W^{\alpha \beta}$ gives the result
\begin{eqnarray}
M_{fi} & = & - \frac{e_a^2 Q g (Q^2 - g^2)}{72 \pi^2 M^4}
\ln \left( \frac{M}{m_a} \right) \nonumber \\
& & \times
[m_a \overline{u} (p_a^\prime) u(p_a) 
\widetilde{F}_{\alpha \beta} (k_3)
F^{\alpha \beta} (k_4)
-i (1/2) \overline{u} (p_a^\prime) \hat{q} \gamma_5 u(p_a)
F_{\alpha \beta} (k_3)
F^{\alpha \beta} (k_4)].
\label{ea17}
\end{eqnarray}

Writing $\widetilde{F}_{\alpha \beta}$ 
and $F_{\alpha \beta}$ in
terms of the components of the electric and magnetic fields
gives
\begin{equation}
\widetilde{F}_{\alpha \beta} (k_3) F^{\alpha \beta} (k_4) =
2[{\bf E} (k_3) \cdot {\bf B} (k_4) + {\bf B} (k_3)
\cdot {\bf E} (k_4)],
\label{ea18}
\end{equation}
\begin{equation}
F_{\alpha \beta} (k_3) F^{\alpha \beta} (k_4) =
2 [{\bf B} (k_3) \cdot {\bf B} (k_4) - {\bf E} (k_3)
\cdot {\bf E} (k_4)].
\label{ea19}
\end{equation}
This allows us to rewrite Eq. (\ref{ea17}) as
\begin{eqnarray}
M_{fi} & = & \frac{e_a^2 Q g (Q^2 - g^2)}{36 \pi^2 M^4}
\ln \left( \frac{M}{m_a} \right) \nonumber \\
& & \times \{-m_a \overline{u} (p_a^\prime) u(p_a)
[{\bf E} (k_3) \cdot {\bf B} (k_4) + {\bf B} (k_3)
\cdot {\bf E} (k_4)] \nonumber \\
& & \mbox{} + i (1/2) \overline{u} (p_a^\prime) 
\hat{q} \gamma_5 u(p_a)
[{\bf B} (k_3) \cdot {\bf B} (k_4) - {\bf E} (k_3)
\cdot {\bf E} (k_4)] \}.
\label{ea51}
\end{eqnarray}

The first term of this equation describes a $T$- and $P$-odd
mixed polarizability effect. 
Writing this term as a potential, $U$
and setting $k_3 = k_4 = 0$ 
(as the $T$- and $P$-odd mixed polarizability
involves homogeneous electric and magnetic fields), we get
\begin{equation}
U = \frac{e_a^2 Q g (Q^2 - g^2)}{18 \pi^2 M^4}
m_a \ln \left( \frac{M}{m_a} \right)  {\bf E} \cdot {\bf B}.
\label{eepola2}
\end{equation}
If we define the $T$- and $P$-odd mixed polarizability by the
equation $U = \Delta E
= - \beta {\bf E} \cdot {\bf B}$,
we then have
\begin{equation}
\beta = - \frac{e_a^2 Q g (Q^2 - g^2)}{18 \pi^2 M^4} 
m_a \ln \left( \frac{M}{m_a} \right).
\label{eepola3}
\end{equation}
The above holds for a particle of charge $e_a$.
To find the dyon induced $T$- and $P$-odd mixed polarizability
for a nucleus with $Z$ protons and $N$ neutrons we
must sum over all the quarks, giving
\begin{equation}
\beta_{Z,N} \approx -(Z + N)
\frac{25 e^2 Q g (Q^2 - g^2)}{162 \pi^2 M^4} 
m_q \ln \left( \frac{M}{m_q} \right),
\label{eezn1}
\end{equation}
where $m_q$ is the constituent quark mass.
We used
$\langle N | \overline{q} q | N \rangle
\approx 5 \overline{N} N$, where $N$ is a
proton or neutron and $q$ is either $u$ or $d$
(see, e.g, \cite{CPvws}).
The second term in Eq. (\ref{ea51}) describes 
the interaction of the
spin of a particle with the gradient of the electromagnetic
field's Lagrangian density
[$\propto \bbox{\sigma} \cdot \bbox{\nabla} ({\bf B}^2
- {\bf E}^2)$ in the non-relativistic limit].

\section{Dyon Induced $T$- and $P$-odd Interaction}
\label{stpoint}
\begin{figure}
\leavevmode
\begin{center}
\begin{tabular}{c}
\epsfxsize=245pt
\epsffile{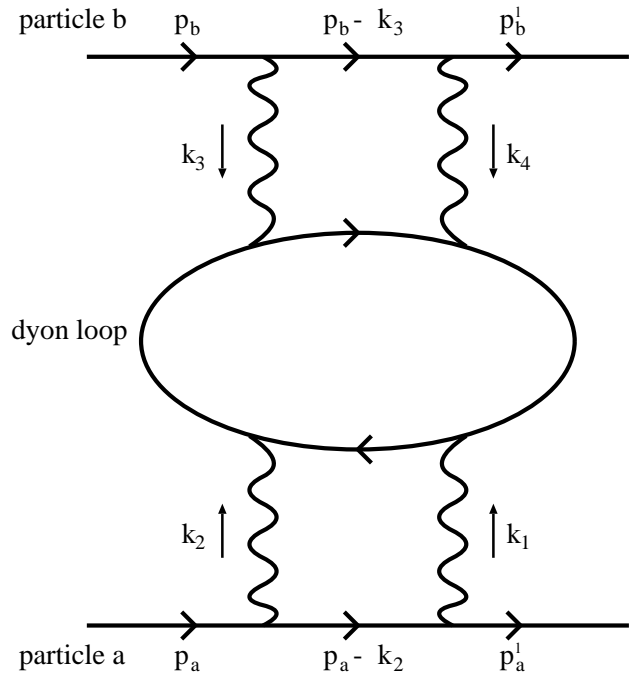}
\end{tabular}
\end{center}
\caption{Diagram showing the dyon induced $T$- and $P$-odd
interaction between two particles.}
\label{f3}
\end{figure}

In this section we will find the $T$- and $P$-odd interaction
between two particles that could be induced by dyons.
We will find the matrix element for the diagram shown
in Fig. \ref{f3}, starting from the matrix element
for Fig. \ref{f2} found above
(\ref{ea17}).
Taking the first term of Eq. (\ref{ea17}) --- we will call it
$M_{fi}^A$ --- and using Eq. (\ref{ea16}) we have
\begin{equation}
M_{fi}^A = - \frac{e_a^2 Q g (Q^2 - g^2)}{36 \pi^2 M^4}
\ln 
\left( \frac{M}{m_a} \right) m_a \overline{u} (p_a^\prime) 
u(p_a)
\varepsilon^{\sigma \beta \omega \alpha} k_{3 \omega}
k_{4 \alpha} A_\sigma (k_3) A_\beta (k_4)
\label{ea40}
\end{equation}
To convert from Fig. \ref{f2} to Fig. \ref{f3}
we take $A_{\sigma} (k_3)$ and $A_\beta (k_4)$
(corresponding to the two free ends) and replace them by
$i e_b^2/(2 \pi)^4 \int D_{\lambda \sigma} (k_3) 
D_{\eta \beta} (k_4)
\overline{u} (p_b^\prime) \gamma^\eta G(p_b - k_3)
\gamma^\lambda u(p_b) \, d^4 k_3$.
Carrying out this 
transformation on Eq. (\ref{ea40}) (and using
$k_4 = -k_3 - q$) gives
\begin{eqnarray}
M_{fi}^{\prime A} & = & -i e_a^2 e_b^2
\frac{Q g (Q^2 - g^2)}{36 \pi^4 M^4}
\ln \left( \frac{M}{m_a} \right) m_a \overline{u} 
(p_a^\prime) u(p_a)
\nonumber \\
& & \times
\varepsilon^{\sigma \beta \omega \alpha} 
\overline{u} (p_b^\prime)
\gamma_\beta
\left \{ \int
\frac{(\hat{p}_b - \hat{k}_3 + m_b)
(-q_\alpha k_{3 \omega} - k_{3 \alpha} k_{3 \omega})
\, d^4 k_3}
{k_3^2 (k_3 + q)^2 [(k_3 - p_b)^2 - m_b^2]} \right \}
\gamma_\sigma u(p_b)
\nonumber \\
& = & -i e_a^2 e_b^2 \frac{Q g (Q^2 - g^2)}{36 \pi^4
M^4} \ln \left( \frac{M}{m_a} \right) m_a \overline{u} 
(p_a^\prime)
u(p_a) \varepsilon^{\sigma \beta \omega \alpha}
\overline{u} (p_b^\prime) 
\gamma_\beta W_{\alpha \omega}^\prime
\gamma_\sigma u(p_b),
\label{ea41}
\end{eqnarray}
(we introduce the dash to distinguish it from $M_{fi}$ for
the previous diagram)
where $W_{\alpha \omega}^\prime$ is like
$W_{\alpha \omega}$ [defined by Eq. (\ref{ea6})], but
with the sign of $q$ changed, and so
we can modify the result (\ref{ea13}) for
$W_{\alpha \omega}^\prime$. 
Because of the $\varepsilon^{\sigma
\beta \omega \alpha}$ only the nonsymmetric part of
$W_{\alpha \omega}^\prime$ will contribute.
So we have
\begin{equation}
M_{fi}^{\prime A} = e_a^2 e_b^2 \frac{Q g (Q^2 - g^2)}
{72 \pi^2 M^4} \ln \left( \frac{M}{m_a} \right)
\ln \left( \frac{M}{m_b} \right) m_a
\overline{u} (p_a^\prime) u(p_a) \varepsilon^{\sigma \beta
\omega \alpha} \overline{u} (p_b^\prime)
q_\alpha \gamma_\beta \gamma_\omega \gamma_\sigma
u(p_b)
\label{ea42}
\end{equation}
Now $\varepsilon^{\sigma \beta \omega \alpha}
q_\alpha \gamma_\beta \gamma_\omega \gamma_\sigma
= 2 i q_\alpha \gamma_5 \sigma^{\sigma \alpha}
\gamma_\sigma
= 6 i \hat{q}
\gamma_5$ and so we have
\begin{equation}
M_{fi}^{\prime A} = i \frac{e_a^2 e_b^2 Q g (Q^2 - g^2)}
{12 \pi^2 M^4} \ln \left( \frac{M}{m_a} \right)
\ln \left( \frac{M}{m_b} \right) m_a
\overline{u} (p_a^\prime) u(p_a) \overline{u} (p_b^\prime) 
\hat{q}
\gamma_5 u(p_b).
\label{ea43}
\end{equation}
A similar process can be carried out for the second term
in Eq. (\ref{ea17}). The total result for the
matrix element of the $T$- and $P$-odd
interaction between particles $a$ and $b$ is
\begin{eqnarray}
M_{fi}^{\prime} & = & i \frac{e_a^2 e_b^2 Q g (Q^2 - g^2)}
{12 \pi^2 M^4} \ln \left( \frac{M}{m_a} \right)
\ln \left( \frac{M}{m_b} \right)
\nonumber \\
& & \times [m_a \overline{u} (p_a^\prime) u(p_a)
\overline{u} (p_b^\prime) \hat{q}
\gamma_5 u(p_b)
+ m_b \overline{u} (p_a^\prime) \hat{q} \gamma_5
u(p_a) \overline{u} (p_b^\prime)
u(p_b)]
\label{ea44}
\end{eqnarray}

\section{Deduced experimental bounds on dyon properties
from electric dipole moments of atoms and the neutron}
\label{sdexp}
In this section we will find limits on the properties of
dyons from experimental bounds on the
electric dipole moments (EDMs) of atoms and the neutron.

\subsection{Dyon induced electron-nucleon interaction}
First we will consider the atomic EDM
that can be produced by the dyon induced $T$- and $P$-odd
electron-nucleon interaction.
To find this interaction we
let particle $a$ be a quark ($q$) and
particle $b$ be an electron ($e$), and sum over the
quarks of the nucleon.

To begin with we will consider the first term of
Eq. (\ref{ea44}) --- the quark scalar-electron pseudoscalar
interaction. It follows from the Dirac equation
and $q = p_e^\prime - p_e$ that
$\overline{u} (p_e^\prime) \hat{q} \gamma_5 u(p_e) =
2 m_e \overline{u} (p_e^\prime) \gamma_5 u(p_e)$.
We rewrite the interaction as a Hamiltonian density:
\begin{equation}
H(x) = -i e_q^2 e^2
\frac{Q g (Q^2 - g^2)}{6 \pi^2 M^4} m_q m_e
\ln \left( \frac{M}{m_q} \right)
\ln \left( \frac{M}{m_e} \right)
\overline{\psi}_{p_q^\prime} \psi_{p_q}
\overline{\psi}_{p_e^\prime} \gamma_5 \psi_{p_e},
\label{e17b1}
\end{equation}
where the wave function, $\psi_p (x)$, is
related to the amplitude $u(p)$ by
$\psi_p (x) = (2 \epsilon_p)^{-1/2} u(p) e^{-i p x}$
($\epsilon_p$ is the energy of the particle).
Summing over the quarks in the
proton and neutron gives
\begin{equation}
H^{sp}_{N} = -i \frac{25 e^4 Q g (Q^2 - g^2)}{54 \pi^2 M^4}
m_q m_e \ln \left( \frac{M}{m_q} \right)
\ln \left( \frac{M}{m_e} \right)
\overline{\phi}_{p_{N}^\prime} (x) \phi_{p_{N}} (x)
\overline{\psi}_{p_e^\prime} (x) \gamma_5 \psi_{p_e} (x),
\label{e17b2}
\end{equation}
where $\phi_{p_N} (x)$ denotes the nucleon
wave functions
and $N$ (nucleon) is either $p$ or $n$.

The Hamiltonian density for the
other term --- the quark pseudoscalar-electron scalar
interaction --- is
\begin{equation}
H(x) = -i e_q^2 e^2
\frac{Q g (Q^2 - g^2)}{12 \pi^2 M^4} m_e
\ln \left( \frac{M}{m_q} \right)
\ln \left( \frac{M}{m_e} \right)
\overline{\psi}_{p_q^\prime} \hat{q} \gamma_5 \psi_{p_q}
\overline{\psi}_{p_e^\prime} \psi_{p_e},
\label{ettc1a}
\end{equation}
Here we have a derivative of a quark axial current
($q_\alpha \overline{\psi}_{p^{\prime}_q} \gamma^\alpha
\gamma_5
\psi_{p_q} = - i \frac{\partial}{\partial x^\alpha}
\overline{\psi}_{p^{\prime}_q} \gamma^\alpha \gamma_5
\psi_{p_q}$),
but we need to express our result in terms of
nucleon wave functions. We make an order of magnitude
estimate using PCAC (partial conservation of axial current)
\cite{CurrAlgebra}:
$\sum_q {e_q}^2 \overline{\psi}_{p^{\prime}_q}
\hat{q} \gamma_5 \psi_{p_q}
\rightarrow (1 \mbox{ GeV}) e^2 
\overline{\phi}_{p^{\prime}_N} \gamma_5 \phi_{p_N}$.
It does not matter exactly what value we choose for the
coefficient on the right hand side of this equation, as due
to the fact that the
main dependence of the interaction on the dyon mass goes
as $1/M^4$, the limit 
on the dyon mass that we will obtain will
not be very sensitive to this choice.
The result for the nucleon pseudoscalar-electron scalar
interaction is
\begin{equation}
H^{ps}_N = i \frac{e^4 Q g (Q^2 - g^2)}{12 \pi^2 M^4}
(1 \mbox{ GeV}) m_e \ln \left( \frac{M}{m_q} \right)
\ln \left( \frac{M}{m_e} \right)
\overline{\phi}_{p_N^\prime} (x) \gamma_5 \phi_{p_N} (x)
\overline{\psi}_{p_e^\prime} (x) \psi_{p_e} (x).
\label{en17a2}
\end{equation}

The interactions in Eqs. (\ref{e17b2}) and
(\ref{en17a2}) can produce atomic EDMs. Note that since
the pseudoscalar $\overline{\psi} \gamma_5 \psi$ is
proportional to the spin of the particle, interaction
(\ref{e17b2}) only has an effect
in atoms with unpaired electrons, while interaction
(\ref{en17a2}) requires an unpaired nucleon.
The size of such scalar-pseudoscalar and pseudoscalar-scalar
interactions can be denoted by the constants
$k_{1N}$ and $k_{3N}$ (see, e.g., \cite{ParNoncons}):
\begin{equation}
H_N^{sp} = i k_{1N} \frac{G}{\sqrt{2}}
\overline{\phi}_{p_{N}^\prime} (x) \phi_{p_{N}} (x)
\overline{\psi}_{p_e^\prime} (x) \gamma_5 \psi_{p_e} (x)
\label{eee1a}
\end{equation}
\begin{equation}
H_N^{ps} = i k_{3N} \frac{G}{\sqrt{2}}
\overline{\phi}_{p_{N}^\prime} (x) \gamma_5 \phi_{p_{N}} (x)
\overline{\psi}_{p_e^\prime} (x) \psi_{p_e} (x).
\label{eee1b}
\end{equation}
Thus, these constants can be written in terms of 
dyon parameters as
\begin{eqnarray}
|k_{1N}| & = & \frac{25 \sqrt{2} e^4}{54 \pi^2 G}
\frac{m_q m_e}{\widetilde{M}^4}
\ln \left( \frac{M}{m_q} \right)
\ln \left( \frac{M}{m_e} \right),
\label{eea2p} \\
|k_{3N}| & = & \frac{\sqrt{2} e^4}{12 \pi^2 G}
\frac{(1 \mbox{ GeV}) m_e}{\widetilde{M}^4} 
\ln \left( \frac{M}{m_q} \right)
\ln \left( \frac{M}{m_e} \right)
\label{eea2r} 
\end{eqnarray}
to order of magnitude accuracy,
where we have defined $\widetilde{M}$ for convenience as
\begin{equation}
\widetilde{M} \equiv \frac{M}{|Q g (Q^2 - g^2)|^{1/4}}.
\label{eea2t}
\end{equation}
From experimental limits on 
atomic EDMs \cite{Commins1994,Hg1995}
and calculations
of the EDMs produced by the interactions (\ref{eee1a})
and (\ref{eee1b}) \cite{ParNoncons,AMMP85}, limits 
on $k_{1N}$ and
$k_{3N}$ can be deduced, and hence limits on 
the dyon parameter
$\widetilde{M}$.
See Table \ref{t1}. Note that in 
Eqs. (\ref{eea2p}--\ref{eea2r}),
in addition to the unknown parameter $\widetilde{M}$ there is
$M$, which is another (independent) unknown parameter, as
$Q$ and $g$ are unknown.
However, since $M$ only appears in logarithms, which are
not very sensitive, we can take
$M = |Q g (Q^2 - g^2)|^{1/4} \widetilde{M} 
\approx 2 \widetilde{M}$,
assuming that $Q \approx e$ and $g \approx 1/(2 e)$
--- this estimate of $g$ is based on the Dirac condition
\cite{Dirac31}: $e g = n / 2$, where $n$ is an integer.
The best bound on $\widetilde{M}$ is
$\widetilde{M} > 6 \mbox{ GeV}$.

Dyons could also induce an 
electron-electron scalar-pseudoscalar
interaction, which could create an atomic EDM in atoms with
unpaired electrons. However, this effect is negligible
compared
to the effect of the nucleon scalar-electron pseudoscalar
interaction. Since the wave function of the outer, unpaired
electron is enhanced at small distances, the main contribution
to the interaction is from $r < a/Z$, where $a$ is the
Bohr radius. In the case of the electron-electron interaction
only the innermost electrons have sizeable wave functions in
this region, while for the nucleon-electron interaction all of
the nucleons are able to contribute, so the electron-nucleon
interaction is larger by a factor $\sim Z + N$. In
addition, the wave function of the outer electron becomes
very large near the nuclear surface and this enhances the
nucleon-electron interaction by a factor $\sim 10$. See
Ref. \cite{SF1978} for a similar result concerning
the relative contributions of weak electron-nucleon
and electron-electron interactions.

\subsection{Dyon induced quark-quark interaction}
Now we consider the $T$- and $P$-odd quark-quark interaction.
This corresponds to Eq. (\ref{ea44}), with particles $a$ and
$b$ both quarks. The interaction between quarks $q_1$ and
$q_2$ is
\begin{equation}
H(x) = i {e_{q_1}}^2 {e_{q_2}}^2
\frac{Q g (Q^2 - g^2)}{6 \pi^2 M^4} m_{q_1}^\prime m_{q_2}
\ln \left( \frac{M}{m_{q_1}} \right)
\ln \left( \frac{M}{m_{q_2}} \right)
\overline{\psi}_{p_{q_1}^\prime} \gamma_5 \psi_{p_{q_1}}
\overline{\psi}_{p_{q_2}^\prime} \psi_{p_{q_2}}
\label{enf2a}
\end{equation}
(of course there is also the scalar-pseudoscalar interaction).
Here we used
$\overline{u} (p_{q_1}^\prime) \hat{q} \gamma_5 u(p_{q_1}) =
-2 m_{q_1}^\prime \overline{u} (p_{q_1}^\prime) \gamma_5 
u(p_{q_1})$,
as we did for the electron pseudoscalar in
Sec. \ref{sdexp}, but note that  $m_{q_1}^\prime$ is the
current quark mass: $\sim 5 \mbox{ MeV}$. The $m_{q_2}$ comes
from $\hat{p}_{q_2} u(p_{q_2}) = m_{q_2} u(p_{q_2})$
(as in the Appendix) and is the constituent quark mass
($\approx m_N / 3$) as it takes into account the effects 
of the
gluon cloud. It is not used in
$\overline{u} (p_{q_1}^\prime) \hat{q} \gamma_5 u(p_{q_1}) =
-2 m_{q_1}^\prime \overline{u} (p_{q_1}^\prime) \gamma_5 
u(p_{q_1})$
as this involves the difference of two Dirac equations and
so the effects of the gluon cloud cancel out.
The general quark-quark pseudoscalar-scalar interaction can be
written in terms of the dimensionless constant $k_s$ as
(see, e.g., Ref. \cite{KK92})
\begin{equation}
H = i  k_s \frac{G}{\sqrt{2}}
\overline{\psi}_{p_{q_1}^\prime} \gamma_5 \psi_{p_{q_1}}
\overline{\psi}_{p_{q_2}^\prime} \psi_{p_{q_2}},
\label{enf2b}
\end{equation}
where
\begin{equation}
k_s = \frac{\sqrt{2} e_{q_1}^2 e_{q_2}^2}{6 \pi^2 G}
\frac{m_{q_1}^{\prime} m_{q_2}}{\widetilde{M}^4}
\ln \left( \frac{M}{m_{q_1}} \right)
\ln \left( \frac{M}{m_{q_2}} \right).
\label{ea2a2}
\end{equation}
The calculation done in Ref. \cite{KK92} allows
us to convert the experimental limit on the atomic EDM
of $^{199}$Hg, $|d_A| < 9.7 \times 10^{-28} e \mbox{ cm}$
\cite{Hg1995}, to a limit on $k_s$:
$|k_s| < 2 \times 10^{-6}$.
The quark-quark pseudoscalar-scalar interaction can induce
a neutron EDM.
The limit on the neutron EDM, $|d_n| < 1.1 \times 10^{-25}
e \mbox{ cm}$ \cite{Altarev96,Altarev92,Smith90} implies the
weaker limit $|k_s| < 3 \times 10^{-5}$ \cite{KK92}.
Using the stronger limit, and taking
$e_{q_1}$ and $e_{q_2}$ as $\sim e/3$, gives
$\widetilde{M} > 1.5 \mbox{ GeV}$.

\section{CP violation in K-meson decays}
\label{skmeson}
It would be good if it were possible to make
a comprehensive theory of $CP$- and $T$-violation
based on dyons that could explain the experimentally
observed $CP$-violation in K-meson decays \cite{CCFT1964}.
Unfortunately, no such theory exists at the moment,
but we are able to make some plausible estimates
of dyon induced $CP$-violation in K-meson decays.
However, we find that the effect induced by dyons
is not large enough to explain the observed level of
$CP$-violation.

Because of electroweak unification, since dyons
may induce $T$- and $P$-odd photon-photon scattering
processes, we would expect them to also induce
W-boson scattering processes. The W-boson
scattering tensor should be similar 
to as in Eq. (\ref{eppst}).
$CP$-violation in K-meson decays
could be caused by an
interaction described by the following Lagrangian \cite{Okun}:
\begin{equation}
L = - G_2 \overline{s} \gamma_\alpha (1 + \gamma_5) d 
\overline{s}
\gamma^\alpha (1 + \gamma_5) d.
\label{elcp}
\end{equation}
$CP$ violation would occur if $G_2$ has a non-zero imaginary
part (the real part of $G_2$ is responsible for 
the mass difference
between $K_S^0$ and $K_L^0$). According to \cite{Okun} the
observed $CP$ violation implies that
\begin{equation}
| \mbox{Im} (G_2) | \sim 3 \times 10^{-3} | \mbox{Re} (G_2) |
\label{elcp2}
\end{equation}
and from \cite{Okun} we have
\begin{equation}
\mbox{Re} (G_2) \approx \frac{1}{16 \pi^2} \sin^2 \theta_c
\cos^2 \theta_c G^2 m_c^2,
\label{elcpa8}
\end{equation}
where $\theta_c$ is the Cabibbo angle, $G$ is the Fermi
weak interaction constant ($\approx 10^{-5}/m_p^2$, where
$m_p$ is the proton mass), and $m_c$ is the mass
of the charm quark.

\begin{figure}
\leavevmode
\begin{center}
\begin{tabular}{c}
\epsfxsize=245pt
\epsffile{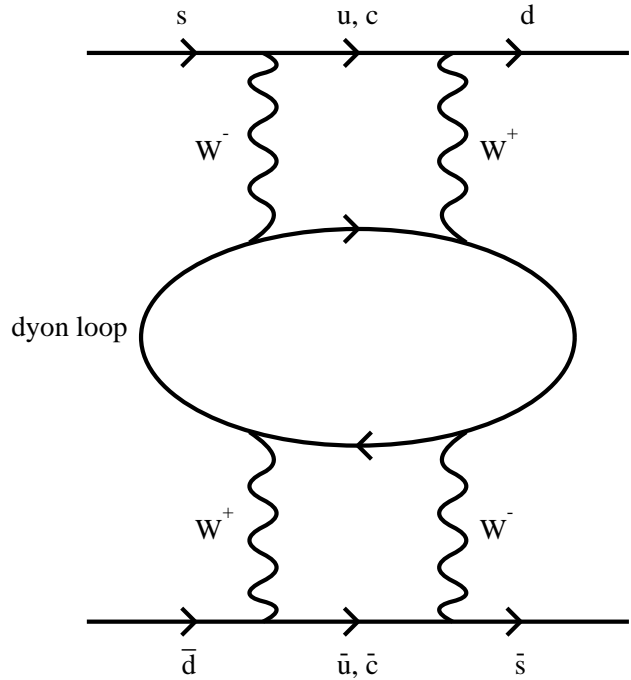}
\end{tabular}
\end{center}
\caption{A diagram through which dyons could induce CP
violation in K-meson decays.}
\label{f4}
\end{figure}

A diagram through which dyons could induce
a CP-violating interaction in K-mesons is shown
in Fig. \ref{f4}. The intermediate quarks may be either
up or charm quarks.
Note that such
a diagram requires two different kinds of dyons, differing in
their electric charges by $e$, to satisfy the conservation
of charge at the $W$-dyon vertices.
(However, even if this is not true the effect may still
appear at higher orders.)

An interaction corresponding
to this diagram is
\begin{equation}
L = i C \overline{d} s \overline{d} s,
\label{ew12a}
\end{equation}
where
\begin{equation}
C \sim \frac{e^4}{96 \pi^2} Q_w g_w (Q_w^2 - g_w^2)
\frac{(m_s^2 - m_d^2) m_c^4}{M_W^4}
\frac{1}{M^4}
\sin^2 \theta_c \cos^2 \theta_c,
\label{ew12b}
\end{equation}
where $m_s$ and $m_d$ are the masses of the strange and
down quarks, $M_W$ is the mass of the W-boson,
and $Q_w$ and $g_w$ are some effective weak electric
and weak magnetic charges for the dyons.
This comes from integrating over the region $M_W <
k < M$, which assumes that the dyon mass is much
greater than the W boson mass. If in fact the dyon mass were
less than the W boson mass then the relevant
region of integration would be $m_c < k < M$ and the
result is
\begin{equation}
C \sim \frac{e^4}{96 \pi^2} Q_w g_w (Q_w^2 - g_w^2)
\frac{(m_s^2 - m_d^2) m_c^4}{M_W^8}
\frac{(M^2 - m_c^2)^2}{M^4}
\sin^2 \theta_c \cos^2 \theta_c.
\label{ew12c}
\end{equation}

We can compare the size of the CP violation induced
by this interaction to the experimentally observed
CP violation in K-meson decays by 
looking at the ratio of the
constant $C$ to $\mbox{Im} (G_2)$ [although
$C$ and $\mbox{Im} (G_2)$ describe two different
types of CP violating interactions, we assume that
the same order of magnitude of either would induce
the experimentally observed CP violation].
If we first consider the case of the dyon mass being
larger than the W boson mass we 
have [using Eqs. (\ref{elcp2}),
(\ref{elcpa8}), and (\ref{ew12b})]
$C / \mbox{Im} (G_2) \sim (M/\mbox{GeV})^{-4}$.
This cannot be larger
than about $10^{-8}$. If we consider
the case of the dyon mass being less than the W boson
mass then we get the result
$C / \mbox{Im} (G_2) \sim 10^{-8} (M^2 - m_c^2)^2 / M^4$,
which has a maximum value of about $10^{-8}$
(here we actually assumed that the dyon mass was larger
than the c quark mass; in the case that it is not we get
a result of similar magnitude). 
These results
suggest that this effect cannot produce
the experimentally observed level of CP-violation 
in K-meson decays.

We can also consider the case when the intermediate quark
is a top quark, rather than up or charm quarks.
If we assume that
$M < M_W$ and integrate
over the region $0 < k < M$ we get the result
\begin{equation}
C \sim \frac{e^4}{1000 \pi^2} Q_w g_w (Q_w^2 - g_w^2)
\frac{(m_s^2 - m_d^2) M^8}{M_W^8 m_t^4} (V_{td} V_{ts})^2,
\label{etsm1}
\end{equation}
where $V_{td}$ and $V_{ts}$ are Cabibbo-Kobayashi-Maskawa
mixing matrix elements. The 
top quark mass is $m_t \approx 170 \mbox{ GeV}$ and,
according to \cite{RevPartPhys}, $|V_{td}|$ and $|V_{ts}|$ 
lie in
the ranges 0.004--0.013 and 0.035--0.042, respectively.
Therefore
$C / \mbox{Im} (G_2)$ has a maximum value of about $10^{-9}$.
Again, this is too
small to give the observed effect.
If we assume that 
$M_W < M < m_t$ and integrate over
the region $M_W < k < M$ we get
\begin{equation}
C \sim \frac{e^4}{96 \pi^2} Q_w g_w (Q_w^2 - g_w^2)
\frac{(m_s^2 - m_d^2)}{m_t^4} 
\frac{(M^2 - M_W^2)^2}{M^4} (V_{td} V_{ts})^2,
\label{eea2at}
\end{equation}
for which $C / \mbox{Im} (G_2)$ has a maximum
value of about $10^{-8}$.
The case of the dyon mass being larger than the
top quark mass gives a result of similar magnitude, with the
maximum value of $C / \mbox{Im} (G_2)$ being
$10^{-8}$.

\begin{figure}
\leavevmode
\begin{center}
\begin{tabular}{c}
\epsfxsize=245pt
\epsffile{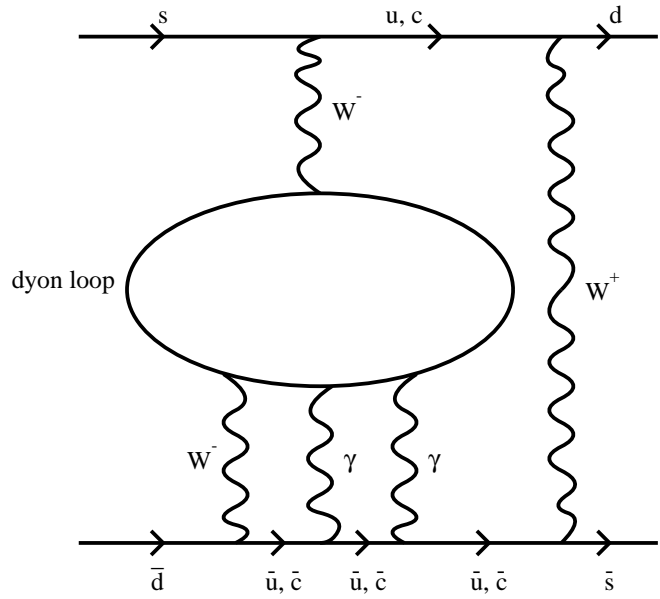}
\end{tabular}
\end{center}
\caption{Another diagram through which dyons could induce CP
violation in K-meson decays.}
\label{f5}
\end{figure}

Dyons could also induce a CP violating interaction through
the diagram shown in Fig. \ref{f5}. However,
once again it seems that it cannot
produce an effect of the
magnitude of the experimentally observed effect.

Finally, the possibility is not excluded that dyons may
generate
an electroweak $\theta$-term ($\propto \theta F_{\mu \nu}
\widetilde{F}^{\mu \nu}$, where $F_{\mu \nu}
= \partial_\mu W_\nu - \partial_\nu W_\mu
+ i g [W_\mu, W_\nu]$) which may give an
effective $CP$-violating interaction
for four W-bosons that does not contain the
fourth power of the dyon mass in the
denominator, and in that case the effect could be
of considerable size even if the dyon mass is large.

\begin{acknowledgements}
This work was supported by the Australian Research Council.
\end{acknowledgements}

\appendix
\section{Example calculation of a contribution to the
matrix element}
\label{sappendix}
In this appendix we calculate the contribution of the term
$c_1 \hat{p}_a g^{\alpha \beta}$ in $W^{\alpha \beta}$
(\ref{ea13}) to $M_{fi}$. We denote this contribution
by $M_{fi}^{(2)}$.

The contribution of the 
1st and 7th terms in Eq. (\ref{ea14}) to $M_{fi}^{(2)}$
(which we will call $M_{fi}^{(2a)}$) is zero, as
they contain the contraction of the
antisymmetric ${\varepsilon_{\alpha \beta}}
^{\mu \nu}$ with $g^{\alpha \beta}$.

The contribution of the 
6th and 12th terms of the equation to
$M_{fi}^{(2)}$ is equal to
(with $c_2 = [i e_a^2 Q g (Q^2 - g^2)]/(360 \pi^4 M^4)$)
\begin{eqnarray}
M_{fi}^{(2b)} & = &
c_1 c_2 A_{\lambda} (k_3) A_{\omega} (k_4) 
k_{3 \rho} k_{4 \sigma}
\varepsilon^{\lambda \omega \rho \sigma}
\overline{u} (p_a^\prime) \gamma_\mu (\hat{p}_a g^{\nu \mu}
- \hat{p}_a g^{\mu \nu} {g^\pi}_\pi) \gamma_\nu u(p_a)
\nonumber \\
& = &
-3 c_1 c_2 A_{\lambda} (k_3) A_{\omega} (k_4) 
k_{3 \rho} k_{4 \sigma}
\varepsilon^{\lambda \omega \rho \sigma}
\overline{u} (p_a^\prime) \gamma_\mu \hat{p}_a 
\gamma^\mu u(p_a)
\nonumber \\
& = & 6 c_1 c_2 A_{\lambda} (k_3) A_{\omega} (k_4) 
k_{3 \rho} k_{4 \sigma}
\varepsilon^{\lambda \omega \rho \sigma} 
m_a \overline{u} (p_a^\prime)
u(p_a),
\label{ea15}
\end{eqnarray}
using $\gamma_\mu \hat{p}_a \gamma^\mu = -2 \hat{p}_a$ and
the Dirac equation: $\hat{p}_a u(p_a) = m_a u(p_a)$.
Now, from the fact that $F_{\mu \nu} (k) =
i (k_\mu A_\nu - k_\nu A_\mu)$ and the definition of the
dual tensor: $\widetilde{F}^{\mu \nu} =
\frac{1}{2} \varepsilon^{\mu \nu \alpha \beta} 
F_{\alpha \beta}$, we
have
\begin{equation}
\varepsilon^{\lambda \omega \rho \sigma} k_{3 \rho}
A_{\lambda} (k_3) k_{4 \sigma} A_\omega (k_4)
= \frac{1}{2} \widetilde{F}^{\sigma \omega} (k_3)
F_{\sigma \omega} (k_4)
\label{ea16}
\end{equation}
Therefore
\begin{equation}
M_{fi}^{(2b)}
= 3 c_1 c_2 m_a \overline{u} (p_a^\prime) u(p_a) 
\widetilde{F}^{\alpha \beta} (k_3)
F_{\alpha \beta} (k_4).
\label{ea30}
\end{equation}

The remaining terms (terms 2--5 and 8--11) can be written as
follows (note that we swap the dummy indices $\mu$ and
$\nu$ around in terms 4, 5, 10, and 11):
\begin{eqnarray}
M_{fi}^{(2c)} & = & c_1 c_2 A_\lambda (k_3)
A_\omega (k_4) k_{3 \rho} k_{4 \sigma}
\overline{u} (p_a^\prime) (\gamma_\mu \hat{p}_a \gamma_\nu
+ \gamma_\nu \hat{p}_a \gamma_\mu) u(p_a)
\nonumber \\
& & \times
(g^{\omega \nu} g^{\alpha \sigma} 
{\varepsilon_\alpha}^{\lambda
\mu \rho}
+ g^{\lambda \nu} g^{\alpha \rho} {\varepsilon_\alpha}^{\omega
\mu \sigma} 
- g^{\nu \sigma} g^{\alpha \omega} 
{\varepsilon_\alpha}^{\lambda
\mu \rho} 
- g^{\rho \nu} g^{\alpha \lambda} 
{\varepsilon_\alpha}^{\omega
\mu \sigma}) 
\nonumber \\
& = & c_1 c_2 A_\lambda (k_3) A_\omega (k_4) 
k_{3 \rho} k_{4 \sigma}
\overline{u} (p_a^\prime) (\gamma_\mu \hat{p}_a \gamma_\nu
+ \gamma_\nu \hat{p}_a \gamma_\mu) u(p_a)
\nonumber \\
& & \times
(g^{\omega \nu} \varepsilon^{\sigma \lambda \mu \rho}
+ g^{\lambda \nu} \varepsilon^{\rho \omega \mu \sigma} 
- g^{\nu \sigma} \varepsilon^{\omega \lambda \mu \rho}
- g^{\rho \nu} \varepsilon^{\lambda \omega \mu \sigma})
\label{ea31}
\end{eqnarray}
Now we have $k_{4 \sigma} A_\omega (k_4) (g^{\omega \nu}
\varepsilon^{\sigma \lambda \mu \rho} - g^{\nu \sigma}
\varepsilon^{\omega \lambda \mu \rho})
= [k_{4 \sigma} A_\omega (k_4) - k_{4 \omega} A_\sigma (k_4)]
g^{\omega \nu} \varepsilon^{\sigma \lambda \mu \rho}
= -i F_{\sigma \omega} (k_4) g^{\omega \nu}
\varepsilon^{\sigma \lambda \mu \rho}
= -i {F_\sigma}^\nu (k_4) \varepsilon^{\sigma \lambda \mu
\rho}$, which corresponds to the 1st and 3rd terms in
Eq. (\ref{ea31}). Doing a similar thing for the 
2nd and 4th terms
as well gives
\begin{eqnarray}
M_{fi}^{(2c)}
& = &
-i c_1 c_2 [k_{3 \rho} A_\lambda (k_3) {F_\sigma}^\nu (k_4)
\varepsilon^{\sigma \lambda \mu \rho}
+ {F_\rho}^\nu (k_3) k_{4 \sigma} A_\omega (k_4)
\varepsilon^{\rho \omega \mu \sigma}]
\overline{u} (p_a^\prime) [\gamma_\mu \hat{p}_a \gamma_\nu
+ \gamma_\nu \hat{p}_a \gamma_\mu] u(p_a)
\nonumber \\
& = &
-c_1 c_2 [\widetilde{F}^{\sigma \mu} (k_3) {F_\sigma}^\nu
(k_4) + {F_\sigma}^\nu (k_3) \widetilde{F}^{\sigma \mu}
(k_4)]
\overline{u} (p_a^\prime) [\gamma_\mu \hat{p}_a \gamma_\nu
+ \gamma_\nu \hat{p}_a \gamma_\mu] u(p_a)
\label{ea32}
\end{eqnarray}
Using the expressions for the $F$ and $\widetilde{F}$
matrices in terms of the electric and magnetic field 
components,
the following identity can be derived:
\begin{eqnarray}
\widetilde{F}^{\sigma \mu} (k_3) {F_\sigma}^\nu (k_4)
+ {F_\sigma}^\nu (k_3) \widetilde{F}^{\sigma \mu} (k_4)
& = & [{\bf E} (k_3) \cdot {\bf B} (k_4) + {\bf B} (k_3)
\cdot {\bf E} (k_4)]
g^{\mu \nu} \nonumber \\
& = & \frac{1}{2} g^{\mu \nu} 
\widetilde{F}^{\alpha \beta} (k_3)
F_{\alpha \beta} (k_4)
\label{ea33}
\end{eqnarray}
Therefore we have
\begin{eqnarray}
M_{fi}^{(2c)} & = &
-c_1 c_2 \widetilde{F}^{\alpha \beta} (k_3) 
F_{\alpha \beta} (k_4)
\overline{u} (p_a^\prime) (\gamma_\mu \hat{p}_a \gamma^\mu) 
u(p_a)
\nonumber \\
& = & 2 c_1 c_2 m_a \overline{u} (p_a^\prime) u(p_a)  
\widetilde{F}^{\alpha \beta} (k_3) F_{\alpha \beta} (k_4)
\label{ea34}
\end{eqnarray}
So, from Eqs. (\ref{ea30}) and (\ref{ea34}), the
contribution of the term $c_1 \hat{p}_a g^{\alpha \beta}$
(in $W^{\alpha \beta}$) to $M_{fi}$ is
\begin{eqnarray}
M_{fi}^{(2)} & = & 
5 c_1 c_2 m_a \overline{u} (p_a^\prime) u(p_a) 
\widetilde{F}^{\alpha \beta}
(k_3) F_{\alpha \beta} (k_4) \nonumber \\
& = & e_a^2 Q g (Q^2 - g^2) / (216 \pi^2 M^4) m_a
\ln (M/m_a) \overline{u} (p_a^\prime) u(p_a) 
\widetilde{F}^{\alpha \beta}
(k_3) F_{\alpha \beta} (k_4) 
\label{ea35}
\end{eqnarray}

\begin{table}
\begin{center}
\begin{tabular}{c c c c c}
Atom & Bound on $|d_A|$ ($e \mbox{ cm}$) & Ref.\ &
Bound on $|k_X|$ & Bound on $\widetilde{M}$
\\
\hline
Tl & $2.9 \times 10^{-24}$&
\cite{Commins1994} & $|k_{1p}| < 1.4 \times 10^{-6}$
& $\widetilde{M} > 6 \mbox{ GeV}$\\
$^{199}$Hg
& $9.7 \times 10^{-28}$&  \cite{Hg1995}
& $|k_{3n}|< 1.6 
\times 10^{-5}$ & $\widetilde{M} > 2.6 \mbox{ GeV}$
\end{tabular}
\end{center}
\caption{Table showing the experimentally determined 
upper bounds on
various atomic EDMs (with references) and
the consequent upper bounds on $k_{1p}$ and $k_{3n}$ and
lower bounds on $\widetilde{M}$.
All bounds are given at the $2$ standard deviation limit.}
\label{t1}
\end{table}

\end{document}